\PassOptionsToPackage{unicode}{hyperref}
\PassOptionsToPackage{hyphens}{url}
\documentclass[
]{article}
\usepackage{amsmath,amssymb}
\usepackage{iftex}
\ifPDFTeX
  \usepackage[T1]{fontenc}
  \usepackage[utf8]{inputenc}
  \usepackage{textcomp} 
\else 
  \usepackage{unicode-math} 
  \defaultfontfeatures{Scale=MatchLowercase}
  \defaultfontfeatures[\rmfamily]{Ligatures=TeX,Scale=1}
\fi
\usepackage{lmodern}
\ifPDFTeX\else
\fi
\IfFileExists{upquote.sty}{\usepackage{upquote}}{}
\IfFileExists{microtype.sty}{
  \usepackage[]{microtype}
  \UseMicrotypeSet[protrusion]{basicmath} 
}{}
\makeatletter
\@ifundefined{KOMAClassName}{
  \IfFileExists{parskip.sty}{%
    \usepackage{parskip}
  }{
    \setlength{\parindent}{0pt}
    \setlength{\parskip}{6pt plus 2pt minus 1pt}}
}{
  \KOMAoptions{parskip=half}}
\makeatother
\usepackage{xcolor}
\usepackage[margin=1in]{geometry}
\usepackage{graphicx}
\makeatletter
\def\maxwidth{\ifdim\Gin@nat@width>\linewidth\linewidth\else\Gin@nat@width\fi}
\def\maxheight{\ifdim\Gin@nat@height>\textheight\textheight\else\Gin@nat@height\fi}
\makeatother
\setkeys{Gin}{width=\maxwidth,height=\maxheight,keepaspectratio}
\makeatletter
\def\fps@figure{htbp}
\makeatother
\providecommand{\tightlist}{%
  \setlength{\itemsep}{0pt}\setlength{\parskip}{0pt}}
\NewDocumentCommand\citeproctext{}{}

\makeatletter
 \let\@cite@ofmt\@firstofone
 \def\@biblabel#1{}
 \def\@cite#1#2{{#1\if@tempswa , #2\fi}}
\makeatother
\newlength{\cslhangindent}
\newlength{\csllabelwidth}
\newenvironment{CSLReferences}[2] 
 {\begin{list}{}{%
  \setlength{\itemindent}{0pt}
  \setlength{\leftmargin}{0pt}
  \setlength{\parsep}{0pt}
  \ifodd #1
   \setlength{\leftmargin}{\cslhangindent}
   \setlength{\itemindent}{-1\cslhangindent}
  \fi
  \setlength{\itemsep}{#2\baselineskip}}}
 {\end{list}}
\usepackage{calc}

\usepackage{listings}

\usepackage{setspace}
\usepackage{footnote}

\usepackage{xcolor}

\newcommand\blfootnote[1]{%
  \begingroup
  \renewcommand\thefootnote{}\footnote{#1}%
  \addtocounter{footnote}{-1}%
  \endgroup
}
\ifLuaTeX
  \usepackage{selnolig}  
\fi
\usepackage{bookmark}
\IfFileExists{xurl.sty}{\usepackage{xurl}}{} 
\hypersetup{
  pdftitle={Identification of distributions for risks based on the first moment and c-statistic},
  pdfauthor={Mohsen Sadatsafavi, Tae Yoon Lee, John Petkau},
  hidelinks,
  pdfcreator={LaTeX via pandoc}}

\title{Identification of distributions for risks based on the first
moment and c-statistic}
\author{Mohsen Sadatsafavi, Tae Yoon Lee, John Petkau}
\date{}

\begin{document}
\maketitle

\blfootnote{From Respiratory Evaluation Sciences Program, Faculty of Pharmaceutical Sciences, the University of British Columbia (MS and TYL); Stanford University School of Medicine (TYL); and Department of Statistics, the University of British Columbia (JP). Corresponece to Mohsen Sadatsafavi (mohsen.sadatsafavi@ubc.ca)}

\subsection{Abstract}\label{abstract}

We show that for any family of distributions with support on {[}0,1{]}
with strictly monotonic cumulative distribution function that has no
jumps and is quantile-identifiable (i.e., any two distinct quantiles
identify the distribution), knowing the first moment and c-statistic is
enough to identify the distribution. The derivations motivate numerical
algorithms for mapping a given pair of expected value and c-statistic to
the parameters of specified two-parameter distributions for
probabilities. We implemented these algorithms in R and in a simulation
study evaluated their numerical accuracy for common families of
distributions for risks (beta, logit-normal, and probit-normal). An area
of application for these developments is in risk prediction modeling
(e.g., sample size calculations and Value of Information analysis),
where one might need to estimate the parameters of the distribution of
predicted risks from the reported summary statistics.

\newpage

\subsection{Background}\label{background}

In risk prediction modeling, one might desire to know the distribution
of risks in a population of interest. For example, contemporary sample
size calculations for external validation of risk prediction models
require knowledge of the distribution of predicted risk (Richard D.
Riley et al. 2021; Pavlou et al. 2021; Richard D. Riley et al. 2024a).
Similarly, the joint distribution of the net benefit of the strategies
involving using or not using the model is required to compute metrics
such as the Expected Value of Perfect Information for risk prediction
models Sadatsafavi et al. (2023). The distribution of predicted risks is
seldom presented in contemporary risk prediction modeling studies. Riley
et. al.~note that sometimes the report on the development or validation
of the model provides some clue, for example, through a histogram of
predicted risks Richard D. Riley et al. (2024b); but parameterizing the
distribution of risks based on scalar performance metrics can be an
objective first step, and at times the only step available to the
investigator.

Arguably, the majority of development or validation studies report two
basic summary statistics: the sample mean of the outcome (or predicted
risks) and the c-statistic of the model summarizing its discriminatory
power. There is partial information in these metrics that can provide
insight into the distribution of predicted risks. Generally speaking,
the sample mean is associated with the location of the central mass of a
distribution, while the c-statistic provides indirect information about
the spread of predicted risks (models with low c-statistic have
concentrated mass, while those with high c-statistic are generally more
variable). Given that common distributions for modeling probabilities
(e.g., beta, or normal distribution for logit-transformed probabilities
{[}aka logit-normal distribution{]}) are indexed by two parameters, it
is intuitive that within a given family of such distributions, knowing
the expected value and c-statistic should uniquely identify the
distribution. But to the best of our knowledge, no rigorous proof has
hitherto been offered.

This works builds on such intuition, and proves the uniqueness of such
identification. We also provide software that encapsulates basic
implementation of such identification, and examine its face validity
through brief simulation studies.

\subsection{Notation and definitions}\label{notation-and-definitions}

Let \(F\) be the strictly monotonic CDF from the family of distributions
of interest with support on \([0,1]\). Let \(\pi\) be a random draw from
this distribution, and \(Y|\pi \sim \mbox{Bernoulli}(\pi)\) the
corresponding `response' variable. Let \(m\) be the first moment of
\(F\), and \(c\) its c-statistic. \(c\) is the probability that a random
draw from \(\pi\)s among `cases' (those with \(Y=1\)) is larger than a
random draw from \(\pi\)s among `controls' (those with \(Y=0\)).
Formally, \(c:=P(\pi_2 > \pi_1 | Y_2=1, Y_1=0)\) with \((\pi_1,Y_1)\)
and \((\pi_2,Y_2)\) being two pairs of predicted risks and observed
responses, with \((\pi_1,Y_1) \mathrel{\perp} (\pi_2,Y_2)\) .

We define \(F\) as \emph{quantile-identifiable} if knowing any pair of
quantiles fully identifies the distribution. We note that common
two-parameter distributions for probabilities, such as beta
(\(\pi\sim \mbox{Beta}(\alpha,\beta)\)), logit-normal
(\(\mbox{logit}(\pi) \sim \mbox{Normal}(\mu,\sigma^2)\), where
\(\mbox{logit}(\pi):=\mbox{log}(\pi/(1-\pi))\)) and probit-normal
(\(\Phi^{-1}(\pi) \sim \mbox{Normal}(\mu,\sigma^2)\) where \(\Phi(x)\)
is the standard normal CDF) satisfy the strict monotonicity and
quantile-identifiability. The quantile-identifiability of the beta
distribution was established by Shih Shih (2015). For the logit-normal
and probit-normal distributions, it is immediately deduced from the
monotonic link to the normal distribution and the
quantile-identifiability of the latter (noting that each quantile
establishes the equality \(\sigma\Phi^{-1}(p_i)+\mu=q_i\), and the
system of two linear equations for \{\(\mu\), \(\sigma\)\} has at most
one solution).

\subsection{Lemma}\label{lemma}

Consider \(F\), a family of probability distributions with support on
{[}0,1{]}, with the following characteristics:

\begin{itemize}
\tightlist
\item
  the CDF is strictly monotonic and has no jumps;
\item
  the distribution is quantile-identifiable.
\end{itemize}

Then \(F\) is uniquely identifiable from \{\(m\), \(c\)\}.

\subsection{Proof}\label{proof}

First, we use the standard result that for \(\pi\) as a non-negative
random variable, \(\mathbb{E}(\pi)=1 - \int_0^1  F(x)dx\). Thus, knowing
\(m\) is equivalent to knowing the area under the CDF. Next, applying
Bayes' rule to the distribution of \(\pi\) among cases (\(P(\pi|Y=1)\))
and controls (\(P(\pi|Y=0)\)) reveals that the former has a PDF of
\(xf(x)/m\) and the latter \((1-x)f(x)/(1-m)\), where \(f(x):=dF(x)/dx\)
is the PDF of \(F\). Thus we have

\begin{flalign*}
m(1-m)c&=\int_0^1\left(\int_0^x(1-y)f(y)dy\right)xf(x)dx \\ &=\int_0^1\left(\int_0^x f(y)dy\right)xf(x)dx-\int_0^1\left(\int_0^x yf(y)dy\right)xf(x)dx \\ &=\int_0^1F(x)xf(x)dx-\int_0^1g(x)G(x)dx,
\end{flalign*}

where \(g(x)=xf(x)\) and \(G(x)=\int_0^x g(y)dy\). Integration by parts
for the first term and a change of variable for the second term result
in

\begin{gather*}
m(1-m)c=\frac{1}{2}xF^2(x)|_0^1-\frac{1}{2}\int_0^1F^2(x)dx-\frac{1}{2}G^2(x)|_0^1=\frac{1}{2}-\frac{1}{2}\int_0^1F^2(x)dx-\frac{1}{2}m^2;
\end{gather*}

i.e., among the subset of \(F\)s with the same \(m\), \(c\) is
monotonically related to \(\int_0^1F^2(x)dx\). As such, the goal is
achieved by showing that \{\(\int_0^1F(x)dx\), \(\int_0^1F^2(x)dx\)\}
uniquely identifies \(F\). We prove this by showing that two different
CDFs \(F_1\) and \(F_2\) from the same family and with the same
\(\int_0^1F(x)dx\) cannot have the same \(\int_0^1F^2(x)dx\).

Given that both CDFs are anchored at (0,0) and (1,1), are strictly
monotonic, and have the same area under the CDF but are not equal at all
points, they must cross. However, they can only cross once, given the
quantile-identifiability requirement (if they cross two or more times,
any pairs of quantiles defined by the crossing points would fail to
identify them uniquely).

Let \(x^*\) be the unique crossing point of the two CDFs, and let
\(y^*=F_1(x^*)=F_2(x^*)\) be the CDF value at this point. We break
\(\int_0^1(F_1^2(x)-F_2^2(x))dx\) into two parts around \(x^*\):

\begin{gather*}
\int_0^1 (F_1^2(x)-F_2^2(x))dx=\int_0^{x^*} \left(F_1(x)-F_2(x)\right)\left(F_1(x)+F_2(x)\right)dx+ \int_{x^*}^1 \left(F_1(x)-F_2(x)\right)\left(F_1(x)+F_2(x)\right)dx.
\end{gather*}

Without loss of generality, assume we label the \(F\)s such that
\(F_1(x) > F_2(x)\) when \(x \in (0,{x^*})\). In this region, due to the
CDFs strictly increasing,
\(0<F_1(x)+F_2(x) < F_1({x^*})+F_2({x^*}) =2y^*\). As such, replacing
\(F_1(x)+F_2(x)\) by the larger positive quantity \(2y^*\) will increase
this term. As well, in the \(x\in ({x^*},1)\) region,
\(F_1(x)-F_2(x) < 0\), and
\(0<F_1({x^*})+F_2({x^*})=2y^*<F_1(x)+F_2(x)\). As such, replacing
\(F_1(x)+F_2(x)\) by the smaller positive quantity \(2y^*\) will also
increase this term. Therefore we have

\begin{gather*}
\int_0^1 (F_1^2(x)-F_2^2(x))dx < 2y^* \left(\int_0^{x^*} (F_1(x)-F_2(x))dx+\int_{x^*}^1 (F_1(x)-F_2(x))dx\right),
\end{gather*}

and the term on the right-hand side is zero because of the equality of
the area under the CDFs. Therefore,
\(\int_0^1 (F_1^2(x)-F_2^2(x))dx < 0\), establishing the desired result.

\subsection{Implementation}\label{implementation}

We have implemented a set of numerical algorithms for finding the
parameters of specified families of distributions of the above-mentioned
class given a known mean and c-statistic in the accompanying
\href{https://cran.r-project.org/package=mcmapper}{\textbf{mcmapper}} R
package Sadatsafavi (n.d.). Our implementation is informed by the above
developments, in particular the two equalities for any distribution with
the required characteristics:

\begin{gather*}
\int_0^1F(x;\lambda)dx = 1-m;\\
\int_0^1F^2(x;\lambda)dx = 1-2cm+(2c-1)m^2,
\end{gather*}

where \(\lambda\) is the set (typically a pair) of parameters indexing
the distribution.

The CDF and its square for the class of distributions that satisfy the
identification requirement are generally well-behaved: they are smooth,
strictly monotonic functions within the unit square. As such, these
integrals can, for the most part, be evaluated using general numerical
integrators. Solving this system of equations can also be programmed as
a two-variable optimization problem, for example via the gradient
descent algorithm that finds the value of \(\lambda\) that minimizes the
quadratic error
(\(\int_0^1F(x;\lambda)dx-[1-m])^2+(\int_0^1F^2(x;\lambda)dx-[1-2cm+(2c-1)m^2])^2\).
The \textbf{mcmap\_generic()} function in the \textbf{mcmapper} package
implements this general algorithm for a general CDF that is indexed by
two parameters. It relies on base R's \textbf{integrate()} function for
computing the two integrals, and base R's \textbf{optim()} function for
the gradient descent component.

This generic mapping algorithm can be improved for specific cases. For
example, for the beta and probit-normal distributions, knowing \(m\)
immediately solves for one of the two distribution parameters. For the
beta distribution, the relationship is \(m=\alpha/(\alpha+\beta)\). As
such, the optimization problem can be reduced to a one-dimensional
root-finding: solve for \(\alpha\) in
\(\int_0^1\mathcal{B}^2(x;\alpha,\alpha\frac{1-m}m)dx = 1-2cm+(2c-1)m^2\),
where \(\mathcal{B}\) is the beta distribution CDF (i.e., incomplete
beta function). The \textbf{mcmap\_beta()} function implements this
approach. For the probit-normal distribution, the relationship is
\(\Phi(\mu/\sqrt(1+\sigma^2))=m\). Again, expressing \(\mu\) as a
function of \(m\) and \(\sigma\) reduces the problem to one-dimensional
root-finding: solve for \(\sigma\) in
\(\int_0^1\Phi^2(\frac{\Phi^{-1}(x)-\Phi^{-1}(m)\sqrt(1+\sigma^2)}{\sigma})dx = 1-2cm+(2c-1)m^2\)
(one can alternatively express \(\sigma\) in terms of \(m\)). The
\textbf{mcmap\_probitnorm()} function implements this algorithm. For the
\mbox{logit}-normal distribution, moments are not analytically
expressible Holmes and Schofield (2022), and our base implementation of
\textbf{mcmap\_\mbox{logit}norm()} uses the gradient descent algorithm
fine-tuned for this particular case (e.g., log-transforming \(\sigma\)
to enable unconstrained optimization). However, this algorithm may fail
to converge for extreme values of \(m\) or \(c\). An alternative
implementation, invoked by default when \(m \notin [0.01, 0.99]\) or
\(c \notin [0.55, 0.95]\), breaks the problem into two nested
one-dimensional root-finding algorithms. It solves for \(\sigma\) in
\(\int_0^1\Phi^2(\frac{\log(x/(1-x))-\mu(m, \sigma)}{\sigma})dx = 1-2cm+(2c-1)m^2\),
where \(\mu(m, \sigma)\) returns the \(\mu\) parameter of the
logitnormal distribution given its mean and \(\sigma\). In turn,
\(\mu()\) solves for \(\mu\) in
\(\int_0^1\Phi(\frac{log(x/(1-x))-\mu}{\sigma})dx=m\). The latter takes
advantage of the infinite series for \(\mu\) derived by Holmes and
Schofield Holmes and Schofield (2022).

\subsection{Brief simulation studies}\label{brief-simulation-studies}

We conducted brief simulation studies to determine the numerical
accuracy of our algorithms in recovering the two parameters of the beta,
logit-normal, and probit-normal distributions. Our approach involved
identifying the parameters of the distribution for a given
\{\(m\),\(c\)\} value, empirically estimating the \{\(m\),\(c\)\} of
this distribution via randomly drawing a large number of risks and
response values, and comparing these estimates with the original values.

We performed this simulation for each combination of
\(m=0.01,0.02,\dots,0.49,0.50\) and \(c=0.51, 0.52, \dots, 0.98,0.99\).
Note that \(m\) was capped at 0.5 because for all three distributions,
the solutions for \{\(m\), \(c\)\} and \{\(1-m\), \(c\)\} are mirror
images of each other around \(m=0.50\) (for the beta distribution, the
two solutions have their \(\alpha\) and \(\beta\) interchanged, and for
the other two, the \(\mu\)s are negative of each other). As such the
solution for \(m>0.5\) can be derived from the corresponding solution
for \(1-m\).

For each \{\(m\),\(c\)\} value, the size of the simulated data was based
on targeting standard error (SE) of 0.001 around both metrics. This
value was chosen so that at least the two digits after the decimal
points remain significant. For \(m\) we used the Wald-type SE formula to
determine the required sample size. For \(c\), we used the SE formula by
Newcombe Newcombe (2006). The target sample size was the maximum of the
two values. We calculated the differences between \{\(m\),\(c\)\} and
\{\(\hat m\), \(\hat c\)\}.

\begin{figure}
\includegraphics[width=1\linewidth]{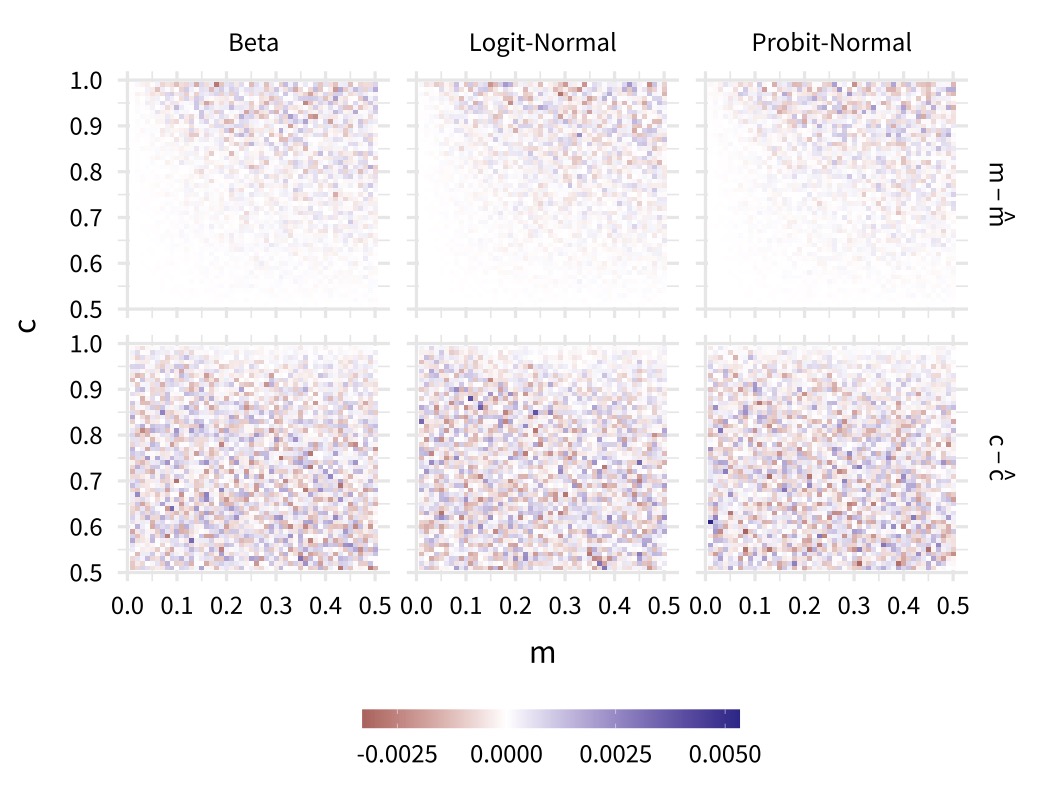} \caption{Difference between original and recovered (via simulation) values of $m$ (top) and $c$ (bottom) for each of the beta (left), logit-normal (middle), and probit-normal (right) distributions}\label{fig:sim-fig}
\end{figure}

Simulation results are shown in Figure \ref{fig:sim-fig}, and illustrate
that our algorithm could recover the parameters successfully for typical
\{\(m\),\(c\)\} values, with error not exceeding 0.005.

\subsection{Remarks}\label{remarks}

In this work, we showed that a family of strictly continuous
distributions with support on {[}0,1{]} can be identified by knowing
their expected value and c-statistic. The regularity condition that we
impose was `quantile identifiability': that knowing two quantiles of the
distribution is sufficient for characterizing them. The general
two-parameter family of distributions that are often used to model
probabilities, including beta, logit-normal, and probit-normal, satisfy
this condition. To test the numerical accuracy of our proposed algorithm
for mapping mean and c-statistic to the parameters of two-parameter
distributions, we performed brief simulation studies. The mapping
algorithm is developed into the publicly available \textbf{mcmapper} R
package.

In our simulation study, we showed satisfactory performance for a
plausible range of mean and c-statistic. However, these algorithms might
struggle at extreme ranges, especially for the c-statistic (e.g., for
\(c<0.51\) or \(c>0.99\)). This is because the CDF can be very flat
(within the floating point precision of the computing device) in
extended parts of its range in such cases, making numerical integration
inaccurate. As well, the surface of the function mapping \(\{m,c\}\) to
the parameters of the distribution might be flat around the solution in
such extreme cases, causing the gradient descent algorithm to fail to
converge. In general, extreme values of the c-statistic are not
plausible in real-world situations. A c-statistic that is very close to
1, for example, might indicate a complete separation of cases and
controls. Reciprocally, a c-statistic that is very close to 0.5 signals
no variability in risks. In such situations, the appropriateness of
modeling the risk as a continuous distribution might be doubtful.

In pursuing this development we were motivated by several recent
developments in risk prediction modeling that require specifying the
distribution of predicted risks in a population, in particular sample
size calculations and Value of Information analyses(Richard D. Riley et
al. 2021; Sadatsafavi et al. 2023). The report on the development of a
risk prediction model often contains an estimate of the expected value
of risk \(\hat{\mathbb{E}}(Y)\) and the c-statistic of the model for
predicting \(Y\). Assuming that the model is moderately calibrated in
the development population (i.e., \(\mathbb{E}(Y|\pi)=\pi\)), the
identifiability results in this work can directly be applied to recover
the parameters of the distribution. In applying these developments, the
onus is on the investigator to decide on the appropriateness of the
family of distributions. Sometimes there is auxiliary information in a
report, such as the shape of the receiver operating characteristic curve
or the histogram of predicted risks, that the investigator can draw on
(e.g., which family of distributions generates the best matching
histogram once the parameters are recovered?), and perform sensitivity
analyses as required.

Do these developments apply with other measures of centrality such as
median and mode? This is obvious for some specific cases. For example,
for both the \mbox{logit}-normal and probit-normal distributions,
knowing the median directly identifies \(\mu\); and \(\sigma\) is then
uniquely determined by \(c\). As well, the mode of the beta distribution
establishes a linear relation between its parameters, so one can express
one parameter in terms of the other given the mode, and solve for the
remaining parameter by knowing \(c\). However, in the Appendix, we
provide counterexamples that show that under the regularity conditions
stated here, the lemma is not generally applicable with the median or
mode.

\section{References}\label{references}

\phantomsection\label{refs}

\clearpage

\section{Appendix}\label{appendix}

To generate counterexamples, we employ the fact that for any
distribution \(f\) with support on {[}0,1{]}, the distribution whose PDF
is the mirror image of \(f\), corresponding to \(\pi \rightarrow 1-\pi\)
transformation, has the same c-statistic as \(f\) (as one can change the
label of cases and controls and arrive at the original distribution).
Our counterexamples involve defining the family of distributions based
on the mixture of triangular distributions whose mirror image is also
the member of the same family. To proceed, define a triangular
distribution by (\(a\), \(b\), \(c\)), where \(a\) is the lower bound,
\(b\) is the upper bound, and \(c\) is the mode of the distribution. For
the case of mode, consider the family of distributions indexed by a
single parameter \(a \in (0,1)\), constructed by a mixture of two
triangular distributions such that 3/4 of the mass is from the
triangular distribution \((0,1,0.5)\) and 1/4 is from \((0,1,a)\). This
family of distributions satisfies the regularity conditions. Direct
evaluation of the PDF of this familty of distributions reveals that the
mode is always at \(\pi=0.5\). A mirror image of any member of this
distribution corresponds to another distribution within the same family
with \(a \rightarrow 1-a\) transformation. But the mirror distribution
has the same \(c\) (as explained above) and the same mode (at
\(\pi=0.5\)), violating the identifiability claim. A similar
counterexample can be constructed for the median: construct a family of
distributions indexed by \(a \in (0,0.5)\) which is based on an equal
mixture of the two triangular distributions \((0,0.5,a)\) and
\((0.5,1,0.5+a)\). This family satisfies the regularity conditions, with
all members having the same median at \(\pi=0.5\). But again, the
transformation \(\pi \rightarrow 1-\pi\) generates another, distinct
member from the same family (corresponding to \(a \rightarrow 0.5-a\)
transformation), with the same median and \(c\).

\end{document}